\documentstyle{article}

\begin{document}

{\bf AN OCTONIONIC GEOMETRIC (BALANCED) STATE SUM MODEL}

\bigskip

by Louis Crane, Math department KSU

\bigskip
{\bf Abstract} We propose a new 4D state sum model, related to the
balanced model, which is constructed using the octonions, or
equivalently, triality. An effective continuum  physical theory constructed from this model
coupled to the balanced model would have a non-vanishing cosmological constant, chiral asymmetry,
and a gauge group related to the octonions.

\bigskip

{\bf 1. Introduction}

\bigskip

In [1], a discrete state sum model was proposed for the quantum theory
of gravity. The model had a very distinct mathematical form, namely it
was a constrained version of the topological state sum of [2] for the
quantum group $U_Q(SO(4)$, where
the constraints had the effect of destroying the topological nature of
the theory. The specific nature of the constraint was to restrict the
spins appearing in the state sum to those representations of
so(4)=so(3)+so(3) which have equal spins for both summands. For this
reason, we called the resulting state sum ``balanced''.

The derivation of the model in [1] began from a description of the
geometry of a Euclidean 4-simplex in terms of the bivectors on its
faces. The constraint on spins is a quantization of the classical
geometric fact that a bivector in $R^4$ is simple if and only if its
self dual and anti self dual parts have equal absolute values. For
this reason we also describe the model of [1] as a ``geometric''
state sum. The significance of this is that terms in this sum can be
interpreted as quantum geometries, and the state sum seems to be
related to the quantization of general relativity in D=4 (see [3,4].)

The construction of [1] is algebraically special insofar as it
utilizes the accidental isomorphism of so(4) with so(3)+ so(3).
It is therefore natural to ask whether any state sums of the `` same
type'' exist, ie. are there any other constrained versions of TQFTs
around with geometric interpretation?

The purpose of this paper is to construct a second geometric state
sum. This sum is related to the TQFT derived from the
representation theory of $U_q(so(8))$ and makes use in an essential
way of the multiplication table of the octonians [5]. For this reason
we refer to it as an octonionic geometric state sum.

The motivation for this construction is to extend the program for
constructing quantum general relativity from a state sum to include
matter fields. As we will explain below, the new model shares the
characteristics of the balanced one which give us some hope of its
usefulness.

Let us warn the reader at the outset that this paper is not self
contained. A familiarity with [1] is an absolute necessity for reading
it.

The geometric interpretation of the state sum we are presenting is a
Kaluza-Klein model with $R^8$ as fiber
over a four dimensional space time. Thus it would have as a
semiclassical limit a Yang-Mills theory with gauge group
SO(8). This is reminiscent of dimensionally reduced supergravity.
If the state sum model of [1] turns out to be a useful model for
quantum gravity, then the model described here may well be an
interesting way of introducing matter fields.

It is therefore natural to wonder how many geometric state sums
actually exist. The construction of [1] and the present construction
both involve special accidents in the theory of Lie algebras and the
existence of division algebras, namely the quaternions and octonions
respectively. This leads one to conjecture that there are very few
geometric state sum models, at least in D=4. If this is so, then
perhaps they are relevant to the problem of finding a unique
fundamental theory. It is worth noting that the geometric models seem
to use the same bits of algebra as supergravity and supersymmetric
Yang-Mills theory, although in a different way. Of course, it would be
much nicer to have a good general definition of ``geometric state
sum'' before attacking such questions.

This paper mostly consists of an exposition of the octonionic
model. We begin by presenting the model of [1] in a new way, then show
how the new approach has an octonionic
version. We close with a few remarks about the physical program of which
this is a part.

\bigskip

{\bf 2.The Balanced Model and the Maximal Torus.}

\bigskip

Let us begin by proving in a new way that simple bivectors in $R^4$
correspond to elements in the Lie algebra so(4) with equal absolute
values for the projections on the two summands. We can use the inner
product on $R^4$ to raise one index of a bivector, changing it into a
skew symmetric matrix, ie an element of so(4). Since every element of
the group SO(4) is conjugate to an element of a maximal torus of the
group, we can restrict attention to elements of the Lie algebra of a
maximal torus, ie. a Cartan subalgebra.

A maximal torus for SO(4) is just the set of rotations around two
orthogonal planes. The corresponding Cartan subalgebra is just $R^2$
as a lie algebra. If we write coordinates for $R^4$ as $x_i$
i==1,2,3,4, then we can pick for a maximal torus all rotations in the
12 and 34 planes. If we introduce the natural basis for the bivectors
on $R^4$ $b_{ij}=x_i \wedge x_j$ , then the bivectors of the form

\bigskip

$ \alpha b_{12} + \beta b_{34} $

\bigskip

are the ones which correspond to the given Cartan subalgebra.

Within this restricted set of bivectors it is easy to see which ones
are simple: either $ \alpha$ or $ \beta $ must be zero.

Now we would like to see how this relates to the decomposition so(4)
=so(3)+so(3).

This is also very straightforward: the summands correspond to the self
dual and anti-self-dual parts. Hence the summands have absolute values
$| \alpha + \beta |$ and $ | \alpha- \beta | $ . It is now an elementary
calculation to show that the simple bivectors correspond to the
elements of the Lie algebra whose self-dual and anti-self dual
projections have equal absolute values.

Now since both simple bivectors and (anti)self-dual Lie algebra
elements are carried into themselves by conjugation, we can see that
our result is true in general.

Furthermore, since irreducible representations of a Lie algebra are described
by a lattice point in a Cartan subalgebra, it is not hard to see that
we can ``quantize'' the self dual bivectors in $R^4$ in terms of a
subcategory of the representations of so(4).

\bigskip

{\bf 3. Regge-Ponzano Kaluza-Klein Theory.}

\bigskip

The idea of the construction of [1] was to describe the geometry of a
4-simplex in Euclidean 4-space by means of the bivectors on its 2d
faces. If we start not with a single 4-simplex but with a triangulated
4-manifold, we end up with a flat metric on each 4-simplex, with
compatibility on lower dimensional faces. This is a familiar technique
which relativists use to produce an approximation to a Riemannian manifold,
known as Regge calculus. 

The octonionic model is also a state sum for a triangulated
4-manifold, but with data associated to a larger group than SO(4). The
geometric situation which motivates it is a fiber bundle over a four
dimensional space-time where the group of the fiber is SO(8).

In a Kaluza-Klein theory, one considers a higher dimensional manifold
which is the total space of a fiber bundle over spacetime. Einstein
metrics of a restricted type are considered, in which the metric along
each fiber is some fixed, symmetric one. The result is that Einstein's
equation for the total space reduces to Einstein's equation on the
base coupled to Yang-Mills. In effect, the off diagonal piece of the
metric (or more precisely, the frame field) on the total space comes
to play the role of a connection on a vector bundle over the base.

What we are proposing here is a discrete analog of a Kaluza-Klein
theory over a 4-dimensional base $B^4$ . We choose a triangulation of the
base manifold, then choose a flat, affine lifting of each 4-simplex to
the total space T of the bundle, identified with $B^4 \times R^8$. This
is equivalent to picking a flat connection on each 4-simplex. Two
liftings related by a translation in $R^8$ would describe equivalent
geometries. The
curvature of the discretized connection so described becomes evident if
we try to lift a path which loops around a 2-simplex (bone) in $B^4$
(translating each flat lifting if necessary so the lifts of subsequent
pieces of
the path meet) and note that we do not return to the same point in T.

Actually, this is only a partial analog of Kaluza-Klein theory, since
the variables correspond to a choice of a connection, but not of a
metric on the base.

As far as we know, no model of this type has appeared anywhere in the
literature. We are not going to investigate this procedure here as a general
approach to investigating Yang-Mills theory. (Presumably it would be
necessary to couple it to a metric on the base.) Our purpose is to use
it to motivate the construction of a geometric state sum, ie. to
rewrite the geometric data in it in such a way as to allow us to
quantize it by a procedure similar to the construction in [1].

In order to do this, we note that each edge in the triangulation of
$B^4$ is being assigned a displacement vector in $R^8$, and each
2-simplex is assigned a simple bivector. Furthermore, since the
lifting of each whole 4-simplex is affine, the sum of the 2 bivectors
on 2 adjacent 2-simplices is embedded in a hyperplane, and hence is
simple.

If we now take the data on 2-simplices and subdivided 3-simplices as
the fundamental data for our model, we can see that we get a set of
simple bivectors in $R^8$ adding in the same pattern as the 4d
bivectors in the model in [1]. (We remind the reader that this paper
makes no pretense of being self contained).

Up to this point we do not have a natural quantization. The astute
reader has probably also noticed that there was really no motivation
up to this point for the choice of the dimension 8 either. The crucial
point in this argument is that simple bivectors in $R^8$ have a special
quantization.

\bigskip

{\bf 4. Triality, Octonians and Simple Bivectors}

\bigskip

Let us attempt to describe simple bivectors in $R^8$ by a similar
process to the one we used in $R^4$. First, we can use the standard
Euclidean metric to identify the bivectors with elements of the Lie
algebra so(8), just as in D=4. Next, we can pick a basis and choose a
standard Cartan subalgebra, which would be identified with the
bivectors of the form 

\bigskip

$ \alpha b_{12} +  \beta b_{34} + \gamma b_{56} + \delta b_{78}$

\bigskip

with the obvious notation.

The simple bivectors in this set are still just the points where three
of the four coefficients are zero.

Now we do not have a decomposition of the Lie algebra so(8). What we
do have is a very special automorphism of the algebra, called
triality.

The useful thing here is that the operation of the triality map can be
written as a linear map from the Cartan subalgebra to itself. In fact,
the natural basis for so(8) can be decomposed into seven copies of the
basis for a Cartan subalgebra, and the action of triality is the same
on each. The only place where this is explicitly written of which we
are aware is [6].

Specifically, the triality transformation can be written as follows:

\bigskip

$F_{12}=1/2(b_{12}-b_{34} -b_{56}-b_{78})$

$F_{34}=1/2(-b_{12}+b_{34} -b_{56}-b_{78})$

$F_{56}=1/2(-b_{12}-b_{34} +b_{56}-b_{78})$

$F_{78}=1/2(-b_{12}-b_{34} -b_{56}+b_{78})$

\bigskip

(The notation here is not identical to [6].)

Where the Fs are a new basis for the Lie algebra with the same
relations as the natural basis associated to the b's.

This transformation is a direct translation of the multiplication
table of the octonians, just as the decomposition of so(4) used above
is essentially the multiplication law of the quaternions.

In the transformed basis, it is easy to see that the simple bivectors
are characterized by the relations:

\bigskip

$|F_{12}|=|F_{34}|=|F_{56}|=|F_{78}|$ .    (1)

\bigskip

Thus we can represent the quantum version of a simple bivector as the
set of representations of so(8) which are sums of highest weight
representations with highest weights on the diagonal of the weight
lattice.

This leads naturally to the suggestion of a state sum on a
triangulated 4-manifold, which is formally very similar to the one in
[1].

We would label each face and each tetrahedron with an irreducible 
representation of $U_q(so(8))$ whose highest weight satisfied (1), 
join the representations into a $15J_q$ symbol as in the model in [2],
multiply together the evaluations of the $15J_q$ symbols for all the
4-simplices in the triangulation, normalize with the product of powers
of quantum dimensions as in [2], and sum over labellings. Unlike the
case in [1], but as in [2], we would have to sum over a basis for the
tensor operators at each trivalent vertex in our diagrams, since the
representation category of so(8) does not have unique tensor operators
like the representation category of so(3).

\bigskip

{\bf 5. Interpretation of the Octonionic Model}

\bigskip

We have shown that a state sum model can be defined with a
mathematical form very similar to the model in [1], which has at least
a plausible link to a geometrical construction. The fact that it is
necessary to use the triality isomorphism to connect the geometrical
condition of simple bivectors to the categorical algebraic form of
restriction to a subcategory is at once interesting and confusing. 
Since triality is an isomorphism between the vector and half spin
representations, (which is, in fact, at the heart of most supergravity
and superstring models), it is tempting to think of the model as a
Kaluza-Klein model in superspace, ie to think of the displacements on
the edges of the triangulation as lying in a chiral fermionic
direction.

At this point, we have no knowledge of the dynamics of the new model,
but given the relationship between the model of [1] with the
Einstein-Hilbert Lagrangian, [3,4] it is at least worth studying.

In order to use this model as part of a unified theory, it will be
necessary to couple it to the balanced model. We have not yet
investigated this.

Nevertheless, it is possible to discern three features which a physical
model derived from the octonionic model would have.

\bigskip

{\bf PREDICTION 1} {\it The cosmological constant is nonzero.}
\bigskip

Rationale: The model is only finite if we use the representations of
$U_q(SO(8))$ for q a root of unity. The value of q is associated with
a value for the cosmological constant in the resulting effective
theory.

Prediction 1 seemed to be an embarassment until recently.
\bigskip

{\bf PREDICTION 2} {\it The theory is not chirally symmetric}

\bigskip

Rationale: unlike the balanced model, the octonionic model is
sensitive to orientation, as in the topological model in [2].

\bigskip

{\bf PREDICTION 3} {\it The effective gauge group of the theory will
be one of the groups with an octonionic construction}

The exact group we obtain will depend on how we couple the balanced
and octonionic models. Probably, some further algebraic coincidences
will appear in this picture.

Given the extreme difficulty of progress along any line to a unified
theory, this direction is worthy of further study.

\bigskip

{\bf 6. Programmatic Remarks}

\bigskip

This paper is part of a  program (the ``categorical program'') which aims at solving the fundamental
problems of theoretical physics by changing the mathematical structure
within which theories are constructed. Instead of using a Lagrangian 
in a smooth manifold, the theory is to be constructed in a
triangulated, or PL manifold, using the structure of a tensor category
to construct a state sum model on the triangulation.

The motivating physical ideas are that physical lengths have a natural
cutoff (justifying the switch to a PL manifold), and that in quantum
theories symmetry is expressed by the presence of a tensor category,
making it the most natural choice for a building block for a theory. 

The mathematical motivation for the approach is a very profound fit
between the axioms of categorical algebra and the rules of
combinatorial topology [8,9,10,11].

For a number of years, this program was confined to the field of
topological quantum field theory, where it had many successes. In
TQFT, the result of a physical calculation is triangulation
independent; as a result of the axioms of tensor categories it is
possible to construct state sum models with this property.

The model of [1] expanded the categorical program beyond the sphere of
TQFT, creating a categorical state sum which seems closely related to
the study of the Ashtekar/loop variables for quantum gravity [12,13,14].

However, the model of [1] is not topological, so we need to confront the
problem of what happens to the model as the triangulation is
refined. In a sense, the old ultraviolet problem recurs in the new
setting.

We are setting forth here a conjecture as to how this problem might
resolve itself, which we call ``the quantum self-censorship
conjecture''.

\bigskip

{\bf QSC CONJECTURE} {\it The state sum model of [1] has a natural
continuation to lorentzian signature. If a physical experiment is
described in this model by choosing fixed labels on the spacelike boundary of a spacially compact PL
manifold with boundary and a fixed triangulation of the initial and
final hypersurfaces, than there exists a finite triangulation T of
the region consistent with the triangulations of the hypersurfaces, with the property that if probability amplitudes are
calculated for any initial and final data (labellings along the
initial and final hypersurfaces) using the state sum on T then if any
refinement of T is used, the same amplitudes will result. }

\bigskip
 For the purposes of this program,it would actually be almost
equally good to replace this conjecture by a weaker conjecture of ``asymptotic quantum
self censorship'' in which the effects suggested in the heuristic
discussion below make the physical predictions of amplitudes converge as the
triangulation became finer.
\bigskip

{\bf Asymptotic QSC CONJECTURE} {\it The state sum model of [1] has a natural
continuation to lorentzian signature. If a physical experiment is
described in this model by choosing fixed labels on the spacelike boundary of a spacially compact PL
manifold with boundary and a fixed triangulation of the initial and
final hypersurfaces, than there exists a finite triangulation T of
the region consistent with the triangulations of the hypersurfaces, with the property that if probability amplitudes are
calculated for any initial and final data (labellings along the
initial and final hypersurfaces) using the state sum on T and then in
all triangulations refining T, the net of values obtained for the
physical quantity will converge. }

\bigskip

If either conjecture is true, then the theory constructed from the
balanced state
sum model is effectively finite, and hence, an interesting candidate
model for the quantum theory of gravity.

The reason for believing that one or the other conjecture might be true is that the
model in [1] is closely related to a topological theory. If we refine
the triangulation on which we calculate the model but allow no
curvature along the new bones, then the constrained result on the
finer triangulation will be exactly equal to the result on the coarser
one because the constraint of flatness makes the theory topological.

The hope is that given fixed labels (=quantum geometry) on the
boundary, any curvature on the new bones will appear to be below the
Planck scale from the boundary, hence will disappear into a black hole
and not affect the results of the calculation of what is to be
observed at the final hypersurface.

Clearly, these are an optimistic conjectures. At least, they have the
advantage of being capable of investigation.

If the QSC conjecture or its asymptotic version were true, then it would
be extremely natural to look for state sum models which extend the
balanced model and share its properties, in order to add matter fields
to the theory. The octonionic model has the
twin virtues of having an extremely similar form to the balanced
model, and having a good deal of kinship with the most interesting
candidates for grand unified theories. It is this combination of
factors which motivated its development. The two properties of the
balanced model which motivate the QSS conjecture, namely the close
relation to a topological theory and the geometric interpretation,
both apply to the octonionic model as well.

\bigskip

{\bf 7. Further Directions}

\bigskip

Although this paper has already gone considerably beyond what can be
precisely done at this point, it is tempting to guess one step
farther. The most symmetric mathematical object which appears in the
algebraic approaches to string theory and supergravity has not yet
appeared in this approach. We are referring to the exceptional Jordan
algebra, or geometrically viewed, the octonionic projective plane [5].

This object has a natural action of $F_4$ or $E_6$, depending exactly
what structure one asks to preserve. Either of these groups are
interesting gauge groups for a grand unified theory. It would be
interesting (and not terribly difficult) to see if a geometrical state
sum can be constructed out of the algebraic accidents relating these
two groups and their close relatives.

\bigskip

{\bf Bibliography}

\bigskip

1. J. W. Barrett and L. Crane, J. Math. Phys. 39, 3296-3302 (1998).

\bigskip

2. L. Crane, L. Kauffman, and D. Yetter, State Sum Invariants of 4-
Manifolds, JKTR, vol. 6, No. 2, 1997, 177-234

\bigskip

3. L. Crane and D. Yetter, On The Classical Limit of the Balanced
State Sum model gr-qc 9712087

\bigskip

4. J. W. Barrett, The Classical Evaluation of Relativistic Spin
Networks, math.QA/9803063

\bigskip

5. F. Gursey and C. H. Tse On The Role of Division, Jordan and Related
Algebras in Particle Physics, World Scientific, 1996

\bigskip

6. H.Freudenthal Oktaven, Ausnahmegruppen, und Octavengeometrie,
Mathematisch Instituut der Rijksuniversiteit te Utrecht 18 may, 1951

\bigskip

7. L. Crane, Clock and Category; Is Quantum Gravity Algebraic,
J. Math. Phys. 36 1995, 6180-6193

\bigskip

8.L. Crane, 2-D Physics and 3-D Topology, Comm. Math. Phys, 135 1991 615-640

\bigskip

9. L. Crane and I. Frenkel, Four Dimensional Topological Quantum Field
theory, Hopf Categories and the canonical Bases, J. Math. Phys., 35,
1994, 5136-5154

\bigskip

10. M. Mackaay, Spherical 2-categories and 4-Manifold Invariants,
q-alg preprint

\bigskip

11. J. Baez and J. Dolan, Higher Dimensional Algebra and Topological
Quantum Field Theory J. Math. Phys. 36 1995 6073-6105

\bigskip

12. M. P. Reisenberger and C. Rovelli, Sum over Surfaces Form of Loop
Quantum Gravity, Phys. Rev. D 56 1997 3490-3508

\bigskip

13.J. C. Baez, Spin Foam Models, gr-qc 9709052

\bigskip

14. L. Smolin Linking TQFT and Nonperturbative Quantum gravity,
J. Math. Phys. 36 1995 6417-6455

\end{document}